\documentclass{phbauth}
\usepackage{graphicx}

\begin{document}

\begin{frontmatter}

\title{Dicke Superradiance in a magnetoplasma}
\author[address1,address3]{Tobias Brandes\thanksref{thank1}},
\author[address2,address3]{Jun'ichi Inoue},
\author[address2,address3]{Akira Shimizu}
\address[address1]{University of Hamburg, 1. Inst. Theor. Physik,
Jungiusstr. 9, D-20355 Hamburg, Germany}
\address[address2]{Department of Basic Science, University of Tokyo, 
Komaba, Meguro-ku, Tokyo 153-8902}
\address[address3]{Core Research for Evolutional
Science and Technology, JST}

\thanks[thank1]{Corresponding author. E-mail: brandes@physnet.uni-hamburg.de}

\begin{abstract}
We present theoretical results for superradiance, i.e. the collective coherent decay
of a radiating system, in a semiconductor heterostructure under a strong
quantizing magnetic field.
We predict a strong peak (`Dicke-peak') in the emission intensity as a function of time, which should
be observable after a short initial excitation of electrons into the conduction band. This peak has a characteristic
dependence on the magnetic field and should be observable on sub-picosecond time scales.
Furthermore, pumping of electrons and holes into the systems at a rate $T$
leads to a novel kind of oscillations with  frequency $\sim \sqrt{T}$ in the limit of the lowest Landau level.
\end{abstract}

\begin{keyword}
Superradiance, Dicke-effect, magnetoplasma, Maxwell-Bloch equations.
\end{keyword}
\end{frontmatter}

\section{Introduction}
Spontaneous emission of photons from  an ensemble of atoms can lead to superradiance.
First proposed by Dicke in 1954 \cite{Dic54}, it took
nearly 20 years for the first experiment \cite{Skretal73} to observe the predicted superradiant emission peak
as a function of time,
its intensity being proportional to the {\em square} of the number of atoms (molecules) of the emitting gas.

Originally a quantum optical paradigm,
the Dicke superradiance effect has become popular again quite recently not only  
in atomic systems (trapped ions \cite{DeVB96}), but also in other electronic systems such as
coupled artificial atoms (quantum dots) \cite{Fujetal98,BK99,BIS98},
coupled quantum wells \cite{Marburg}, or localized states  in semiconductors
\cite{Shahbazyan}.

In this paper, we discuss the superradiance effect in the electron-hole gas of a two-dimensional semiconductor
quantum well under a strong, quantizing magnetic field.
Belyanin et al. \cite{BKK92} already found the possibility of superradiance in {\em bulk} (3d) semiconductors
in magnetic fields (neglecting interactions among the electrons and without pumping). 
In a recent calculation \cite{BIS98}, we predicted a novel form of superradiance for an optical active region
pumped externally by electron (hole) reservoirs.
In the limit of only the lowest Landau level occupied,  the coherent decay of electron-hole pairs leads to
a peak of the emitted light  with a strong intensity that, as a function of time, shows oscillations with a frequency
$
\omega \simeq \sqrt{2\Gamma T},
$
where $\Gamma $ is the spontaneous decay rate of a single pair and $T$ the rate at which electrons
are pumped in the conduction and holes into the valence band.

\section{Equations of motion}
We start from an eigen basis of wave functions $\phi$,
\begin{equation}
\label{wavefunction}
\phi_{lnk}({\bf x})=\frac{e^{iky}}{\sqrt{L}}\phi_n(x+k/eB)\chi_l(z),
\end{equation}
where $k$ is the momentum in $y$-direction, $B$ the magnetic field in $z$-direction, $\phi_n$ the
$n$-th harmonic oscillator wave function, and $\chi_l$ a standing wave for the lowest subband of the
quantum well. The index $l=c,v$ denotes conduction and valence band, respectively.
Note that the harmonic oscillator wave functions
$\phi_n$ do not depend on the effective electron mass in band $l$.

We start from the equations of motion for the one-particle quantities 
\begin{equation}
\label{p}
p^{ll'}_{nn'kk'}(t):=\langle c^{\dagger}_{nkl}c^{\phantom{\dagger}}_{n'k'l'} \rangle_t.
\end{equation}
Factorizing higher 
order correlation functions, we consider only terms in Eq.(\ref{p}) diagonal
in the momentum $k$, i.e.
$
p^{ll'}_{nn'kk'}(t)=:$ $\delta_{kk'}\delta_{nn'}p^{ll'}_{n}(t)$,
where we neglect scattering within the same band $l$ and consider only $s$-wave scattering in 
Hartree-Fock approximation.

The derivation of the combined Maxwell-semiconductor Bloch equations is done in analogy to
\cite{Benedict}. 
The electric field is assumed to be polarized
in the $x$-$y$ plane and to be extended homogeneously over the optically active volume $L^2L_z$,
where $L_z$ is an estimate for the thickness of the quantum well with area $L^2$.

The average polarization ${\bf P}(t)$ can be written as
\begin{equation}
{\bf P}(t)=\frac{\Phi}{\Phi_0}\frac{{\bf d}}{L^2L_z}
\sum_n[p^{cv}_n(t)+p^{vc}_n(t)],
\end{equation}
where ${\bf d}$ is the dipole moment,
$\Phi=BL^2$ the magnetic flux and $\Phi_0=hc/e$ the flux quantum. For simplicity, we here assume
completly filled  Landau bands, i.e. integer filling factor, and complete spin polarization.

Superradiant solutions of the equations that show the Dicke effect  exist if emitted photons 
escape at a rate $\kappa$, which is introduced as a damping term. This allows us to eliminate
the electric field, and one obtains a closed set of equations
for the polarization $R_n(t)=2p^{vc}_n(t)e^{i\omega t}$ and the inversion $z_n(t):=p^{cc}_n-p^{vv}_n(t)$,
where $\omega$ is the frequency of the electric field in resonance with the transition frequency $\Delta$
between the lowest Landau levels $n=0$ in the valence and conduction band. They read
\begin{eqnarray}
\label{zr}
\frac{\partial}{\partial t}z_n &=&-\frac{\Omega_{\Delta}^2}{\kappa}\Re\left[
R_n^*\sum_nR_n\right]\nonumber\\
&-&\Im \left[\sum_{n'}\gamma_{nn'}R_{n'}R^*_n\right]+T_n\nonumber\\
\frac{\partial}{\partial t}R_n &=&-i\overline{\omega}_nR_n
+\frac{\Omega_{\Delta}^2}{\kappa}z_n\sum_nR_n\nonumber\\
&+& \sum_{n'}\frac{\gamma_{nn'}}{2i}\left[R_{n'}z_n-R_nz_{n'}\right].
\end{eqnarray}
Here, $\overline{\omega}_n=neB/cm_r$, $1/m_r=
1/m_c+1/|m_v|$, and
\begin{equation}
\Omega_{\Delta}=\left(2\pi\Delta d^2
\frac{\Phi}{\Phi_0L^2L_z}\right)^{1/2}.
\end{equation}
Furthermore,
\begin{equation}
\gamma_{nn'}=\frac{1}{L^2}\sum_{\bf q}\tilde{U}({\bf q})|M^{nn'}_{q_y0}(q_x)|^2
\end{equation}
contains the Coulomb interaction $\tilde{U}$ and the electron density matrix elements. 
The term $\partial_t z_n\sim T_n$ has
been derived microscopically \cite{BIS98} and 
simulates pumping of electrons and holes
into the optical active region if the system is open.
\section{Discussion}
\begin{figure}[t]
\begin{center}\leavevmode
\includegraphics[width=1.0\linewidth]{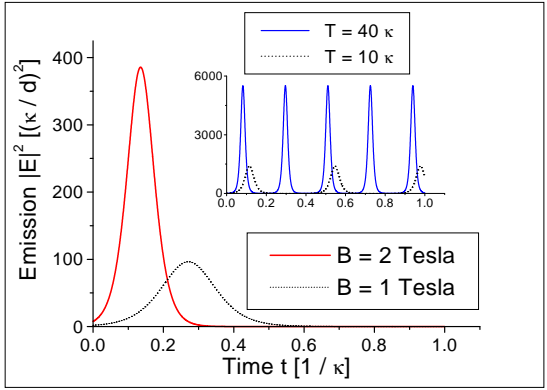}
\caption{ \label{n0}
Time evolution of the emission intensity 
for lowest Landau level. Inset: same with
pumping T at $B=2$ Tesla.}
\label{figurename}\end{center}\end{figure}

In the following, we concentrate on the case of filling factor $\nu=1$; the
case of higher Landau levels will be discussed elsewhere. 
If only the lowest Landau level $n=0$ is involved,
Eq. (\ref{zr}) simplify to two coupled equations for $Z=z_{n=0}$ and $R=R_{n=0}$, where $R$
can be chosen real. The constant of motion
$
J^2:=R^2(t)+Z^2(t)
$
in the case without pumping $T=T_{n=0}=0$ is the total pseudo spin in the Dicke model.
For non-vanishing $T$, it is no longer conserved and one has instead 
\begin{eqnarray}
\label{zj}
\dot{Z}&=&-\frac{\Omega_{\Delta}^2}{\kappa}\left(J^2-Z^2\right)+T\nonumber\\
\dot{J}&=&TZ/J.
\end{eqnarray}
These equation have been derived and discussed previously in the context
of pumped superradiance \cite{BIS98}. In the case $T=0$, they describe
Dicke superradiance, i.e. a strong peak in the emission
$
|E(t)|^2=\left(\Omega_{\Delta}R(t)/d\kappa\right)^2,
$
which is reached for $Z(t)=0$ with a maximum
$\sim (\Omega_{\Delta}^2/d\kappa)^2$. Thus, the maximal emission is proportional
to the square of the number of flux quanta per optical active volume.
Results for the lowest Landau level and typical GaAs/AlGaAs parameters
are shown in Fig.(\ref{n0}). The system starts from an initially
completly inverted state $Z(t=0)$ with a finite polarization $\Re R(t=0)=\Im R(t=0)=0.1$. The time evolution
of the emitted light is in the form of a peak with a maximum $\sim B^2$. This is the so-called Dicke peak:
the radiation is due to transitions of electrons in the conduction band to empty states in the valence band.

The strong magnetic field quenches the kinetic energy so that all
radiators have the same energy difference $\omega$.
They decay not individually, but in a collective way coupled through the
common radiation field. If electrons are pumped into the conduction band and holes into the valence band, the emission
begins to oscillate because after each collective decay the system is `reloaded' again. 
For the parameters used
here, typical time scales are in the sub-picosecond regime.

The microscopic derivation above shows that Dicke superradiance is possible at least
in the quantum limit of only the lowest Landau levels occupied (strong magnetic fields). We have
not included the full magneto-exciton effect into our calculation, but rather considered the case of a
magnetoplasma, where excitonic effects are assumed to be less important. 
\begin{ack}
This work was supported by
CREST, JST, and DFG project Kr 627/9-1
and BR 1528/3-1.

\end{ack}



\end{document}